\documentclass[aps,prb,twocolumn,groupedaddress,floats,showpacs]{revtex4}

\usepackage{bm}

\usepackage{amssymb,amsmath}
\usepackage{latexsym}
\usepackage{graphicx}
\usepackage{epsfig}

\begin{document}

\preprint{APS/123-QED}

\title{ Magnetization Relaxation and Collective Spin Excitations in 
Correlated Double--Exchange Ferromagnets}

\author{M. D. Kapetanakis and I. E. Perakis}

\affiliation{Department of Physics, University of Crete, 
and 
Institute of Electronic Structure \& Laser, Foundation
for Research and Technology-Hellas, Heraklion, Crete, Greece} 

\date{\today}

\begin{abstract}
We study spin relaxation and 
dynamics  of collective spin excitations 
in correlated double--exchange ferromagnets. For this, 
we introduce 
an expansion 
of the  Green's functions equations of motion
that  treats non--perturbativerly 
all correlations between a given number of 
spin and charge excitations and becomes 
exact within a sub--space of states.
Our method 
treats relaxation beyond Fermi's Golden Rule 
while recovering previous variational results
for the spin--wave dispersion.
We find that the momentum dependence 
of the spin--wave dephasing rate 
changes qualitatively due to the on--site Coulomb interaction, 
in a way that resembles experiment,    
and depends on its interplay
with the  magnetic exchange interaction  
and itinerant  spin lifetime.  
We  show that the collective  spin relaxation and its dependence on 
the carrier concentration 
depends sensitively on 
three--body correlations between a spin excitation and a 
Fermi sea electron and hole. 
The above spin dynamics
can be controlled via the itinerant carrier population.

\end{abstract}


\maketitle

\section{Introduction}\label{intro} 
Long--range ferromagnetic order mediated by interactions between 
itinerant 
and localized spins is established  in many different 
materials.\cite{nagaev,tokura,tokura1,III-V}
The manganese oxides 
$\rm{R_{1-x}A_{x}MnO_{3}}$ 
(R=La,Pr,Nd,Sm,$\dots$ and $\rm{A=Ca,Ba,Sr,Pd,}\dots$)
are one prominent example.  
\cite{tokura}
The magnetic and 
transport properties of such itinerant magnetic systems 
are intimately related  and, unlike in 
other ferromagnets, 
can potentially  be  controlled by tuning parameters  
like  the itinerant  carrier density.

Learning how to control  
magnetization dynamics and relaxation 
is important 
for  spintronics applications. \cite{tokura1,wolf,LLG-rev}
One of the challenges facing future 
magnetic devices and memories concerns their speed,  
which is governed by  the 
dynamics 
of the collective spin. 
For small deviations from equilibrium, 
the dynamical magnetic  properties are
determined by the 
spin susceptibility.
\cite{LLG-rev,heinr,mitchell,sinova,halp}
Within the Random Phase approximation (RPA), 
\cite{heinr,furukawa,RPA,nagaev-HP,golosov,sun,papanico} 
which gives  the  spin susceptibility to $O(1/S)$
($S$ is the local spin magnitude), 
magnetization relaxation arises from the interplay between the 
dephasing of the itinerant carrier spin and the magnetic 
exchange interaction. 
 \cite{mitchell,sinova,halp,chovan}
Spin relaxation also arises from inelastic scattering 
processes, 
such as magnon scattering with 
charge excitations.
\cite{kats,golosov,chubukov}
Experimental probes of such effects include neutron scattering,
ferromagnetic resonance, and 
ultrafast magneto--optical pump--probe spectroscopy.
\cite{bigot-05,kampen,kimel-anis,tolk,lupke,kimel-rev,wang-rev,shah,chovan}
The interpretation of  such  experiments 
requires the development 
of  many--body theories of spin dynamics and relaxation.

Our goal in this paper is to develop 
a theory  that describes the  
 local spin Green's function
\begin{equation}
\label{green-S}
\ll 
S^+_{-{\bf Q}} \gg
=-i\theta(t)<
[S^+_{-{\bf Q}}(t),S_{\bf Q}^{-}(0)]>,
\end{equation}
which determines the 
transverse  spin 
susceptibility.
In the above we introduced the collective  spin  operators 
${\bf S}^n_{{\bf q}}
=1/\sqrt{N}
\sum_{j} {\bf S}^n_{j}\rm{e}^{-i {\bf q}\cdot {\bf R}_{j}}$, $n$=x,y,z,   
where  ${\bf S}_{j}$ describe spins localized 
on  $N$ lattice sites at positions  ${\bf R}_{j}$.
$S^{\pm}=S^x \pm i S^y$ are the spin raising/lowering operators and  
$\langle \cdots \rangle$ denotes the grand canonical ensemble
average. 
We consider a model 
Hamiltonian  that accounts for  the most important 
features common in a wide range of
different itinerant ferromagnets, 
$H =K+H_{\rm{exch}}+H_{\rm{AF}}+H_{U}$. 
$K=\sum_{\bf k\sigma} 
\varepsilon_{\bf k} c_{\bf k\sigma}^{\dagger}c_{\bf k\sigma}$
describes a  band 
of itinerant carriers, which in 
the manganites arises from the Mn  
 d--states with $e_{g}$
symmetry. $c^\dag_{\bf k\sigma}$ creates an electron with momentum 
${\bf k}$, spin $\sigma$,
and energy $\varepsilon_{\bf k}$. 
To simplify the 
calculation of correlation effects common 
in many different physical systems, we consider 
a one--band model of $n=1-x$ itinerant electrons
per Mn atom that
hop between nearest--neighbour ltice sites.
The electron concentration is  described by the 
filling factor $n=N_e/N$,
where $N_e$ is the total number of electrons,
which varies from 
0 to 1. 
$\varepsilon_{\bf k}=-t \gamma_{\bf k}$, where 
$\gamma_{{\bf k}}= 2
\sum_{i=1}^{d}\rm{cos}(k_{i}a)$. 
$d$ 
is the system dimensionality and 
$a$ the lattice constant  ($a=\hbar=1$ from now on).
In the manganites, 
0.5$\leq n \le$0.8 in the metallic ferromagnetic regime 
of interest here.  
Our calculation can be extended 
to include the  bandstructure
of individual materials.

In momentum space, the magnetic exchange interaction 
between the local and itinerant spins is
given by 
\begin{eqnarray}
\label{Hexch}
H_{\rm{exch}} &=&-\frac{J}{2\sqrt{N}}\sum_{\bf k,q\sigma}\sigma 
S_{\bf q}^{z}c_{\bf k-q\sigma}^{\dagger}c_{\bf k\sigma}\nonumber\\
&&-\frac{J}{2\sqrt{N}}\sum_{\bf k,q}
\Big(S_{\bf q}^{-}c_{\bf k-q\uparrow}^{\dagger}
c_{\bf k\downarrow}+\rm{H.c}\Big)
\end{eqnarray}
where $\sigma=\pm 1$.  
In the manganites, Eq.(\ref{Hexch}) 
describes the 
Hund's rule onsite interaction between 
the $e_{g}$ carrier spin and the $S=3/2$ local 
magnetic moment of 
the three electrons in the tightly bound $t_{2g}$ Mn orbitals.  
 $J\sim 2 \rm{eV}$ 
and $0.2 \rm{eV}\leq t\leq 0.5 \rm{eV}$
are typical values 
quoted in the  literature, which  give 
$4\leq J/t\leq 10$.\cite{dagotto} 
The Hamiltonian 
$K+H_{\rm{exch}}$ defines the simple double--exchange model.\cite{de,dagotto} 

Here we add to the above minimal double exchange Hamiltonian   
two ubiquitous interactions;
\begin{equation}\label{HU}
H_{U}=
\frac{U}{N}\sum_{\bf kk'q}
c_{{\bf k +q} \uparrow}^{\dagger}
c_{\bf k^\prime- q\downarrow}^{\dagger}
c_{\bf k^\prime \downarrow} c_{\bf k\uparrow}, 
\end{equation}
is the on--site 
(Hubbard) Coulomb 
repulsion.  The Coulomb energy $U\sim 3.5-8 \rm{eV}$
is the largest energy scale in the 
manganites.\cite{dagotto}
This Hubbard interaction is generally hard to treat and its effects on 
the spin dynamics have received less attention.\cite{golosov,sun,kapet2}
\begin{eqnarray}\label{HAF}
H_{\rm{AF}}&=&J_{\rm{AF}}\sum_{\bf k}\gamma_{\bf k}S_{\bf k}^{z}S_{-\bf k}^{z}
\nonumber\\
&&+\frac{J_{\rm{AF}}}{2}\sum_{k}\gamma_{\bf k}
(S_{\bf k}^{+}S_{-\bf k}^{-}+S_{\bf k}^{-}S_{-\bf k}^{+})
\end{eqnarray} 
is the 
direct  super--exchange anti--ferromagnetic 
interaction between the nearest--neighbor 
local spins.  
$J_{\rm{AF}}\sim$0.01t is weak in the manganites.\cite{dagotto} 

Given the large values of $J/t$  in most  systems of interest,
many theories start from the 
strong coupling limit ($J\rightarrow\infty$) of the above Hamiltonian.   
In this limit,  
the itinerant carriers 
can hop on a site only if  
their spin is parallel to the local spin there. 
The kinetic energy 
is  reduced when all  spins are  parallel
(double exchange mechanism \cite{de}), 
which favors the fully polarized half--metallic state 
$|F\rangle$. 
In the classical limit, $S \rightarrow \infty$, 
the problem can  be 
mapped to an effective 
nearest neighbor Heisenberg model with ferromagnetic interactions.
The lowest order ($O(1/S)$) quantum corrections are described by the 
RPA, which for strong couplings 
gives a dispersion 
that again 
coincides with that 
of the nearest neighbor Heisenberg ferromagnet.
\cite{furukawa} 
We may therefore assess the importance of
correlations and quantum fluctuations beyond $O(1/S)$ 
by fitting 
the  Heisenberg 
dispersion to the  experimental 
result 
and looking for deviations.
 
For electron concentrations  $n\ge 0.7$, 
 initial measurements found 
nearest neighbor Heisenberg model spin dynamics.\cite{sw-exp-1}
However, later experiments 
reported  deviations that increase strongly for $n \le$0.7. 
\cite{soft-exp-1,soft-exp-2,soft-exp-3,soft-exp-4,endoh,soft-2D-1,soft-2D-2,ye}
The  Heisenberg model with  nearest neigbor interactions, $J_1$, 
misses a pronounced softening 
near the  Brillouin zone boundary.
This softening is accompanied by 
a strong  increase in the spin--wave damping 
as we approach the zone boundary.
The experimental dispersion 
could be fitted by 
adding a 
fourth--nearest--neighbour ferromagnetic exchange interaction, $J_4$,
to $J_1$ while keeping  $J_3$=$J_2$=0.\cite{ye} 

The above experimental observations
 reveal a new spin dynamics and nonlocal 
correlations that are 
not captured by the strong coupling limit 
of the double exchange model 
(which 
favors local correlations). 
Several scenarios  have been put forward.
 The proposed mechanisms involve, among others, 
magnon  scattering 
with
orbital degrees of freedom,\cite{khal,endoh}
Fermi sea pairs,\cite{kapet1,kapet2} or 
phonons \cite{soft-exp-2,endoh,edwards},  
disorder effects,\cite{disord} bandstructure
effects,\cite{solovyev}
Hubbard interactions \cite{golosov} 
and correlations \cite{kapet2}, 
and the energetic overlap between spin--wave modes 
and the Stoner continuum.\cite{kaplan}
The observed pronounced 
dependence of the spin--wave dynamics on the 
 carrier  concentration 
 puts stringent conditions 
on the theory. 
Ye {\em et.al.} \cite{ye}  
argued that none of the  mechanisms  proposed
so far can fully account for all aspects of this spin dynamics.

The purpose of this paper is 
two--fold. 
First, we study the momentum dependence of the 
spin--wave dephasing rate, with the  focus on 
the role of 
correlations 
between the spin and charge excitations 
and on their interplay with the itinerant  spin dephasing. 
We demonstrate that both the magnitude
and the momentum dependence 
of the spin relaxation  rates 
 depend sensitively on the carrier 
concentration and on   
correlations due to both $J$ and $U$. 
We show that the on--site Coulomb repulsion $U$ 
changes the 
momentum dependence of the spin--wave dephasing rate in a qualitative 
way that resembles the experimental results. 
We also show that our results depend on the interplay 
of $U$ with the itinerant carrier 
spin lifetime, which is finite 
in some systems 
due to interactions 
not included in our Hamiltonian.
\cite{heinr,mitchell}
We compare with the 1/S 
expansion and other approximations and 
find that three--body correlations between spin and 
electron--hole pair excitations play an important role.
Finally, we show 
that the  magnetization relaxation 
can be 
controlled  by tuning the carrier density.
We obtain changes in  the spin relaxation with $n$ that  correlate 
with corresponding changes in the spin--wave 
softening and non--Heisenberg 
behavior.

Second, we 
develop and test a general 
method for  describing  
spin dynamics. 
For this 
we use  a 
truncation scheme 
of the infinite hierarchy of 
Green's function equations of motion
based on an expansion in terms of  correlations.
Our scheme  treats the full dynamics 
induced by the correlations between a
given number of elementary excitations. 
Here we describe all
correlations 
between 
a  local  or  carrier spin excitation 
and an electron--hole  Fermi sea pair and 
 obtain  the exact solution within 
the sub--space of states with up to one Fermi sea  excitation. 
Similar to 
Refs.[\onlinecite{kapet1,kapet2}], 
our method becomes exact in the limits of one electron
($N_e$=1, $n$=1/$N$),   
half filling ($N_e$=$N$, $n$=1), 
and in the atomic limit ($t$=0 for any $n$). It interpolates between 
the weak and strong coupling limits and  
agrees with  exact diagonalization 
results 
for the spin--wave dispersion.\cite{exact}  
Finally, it 
retains
its variational nature
in the limit of zero relaxation rates,
which provides a rigorous bound for the 
spin--wave softening and ferromagnetic phase boundary.
Our approach, 
used before in the context of the 
Hubbard Hamiltonian, \cite{igar,rucken}  
is in the same spirit as the projection and factorization scheme 
 of Ref.[\onlinecite{perakis}], 
used to calculate 
the ultrafast nonlinear optical response 
of systems with a strongly correlated ground state, and the 
 correlation expansion  of Ref.[\onlinecite{corr-exp}].  
It may be extended 
to study  spin correlations 
in non--equilibrium systems
\cite{chovan,shah}
and  the magnetization dynamics of  (III,Mn)V semiconductors.\cite{unpubl} 

The rest of this paper is organized as follows. 
In Section \ref{formul} we discuss the Green's function 
truncation scheme and  
derive a closed system of equations of motion that determine 
 the  spin Green's function. 
In Section \ref{spin-self} we  obtain the spin self--energy
and separate the RPA contribution from the contributions of 
correlations due to  $J$ and $U$.
In Section \ref{num} we discuss our numerical results for the 
spin--wave dephasing  rate and dispersion and compare different approximations.
In Section \ref{num-de} we consider the minimal double exchange model with 
$U$=$J_{AF}$=0, while in 
Section \ref{num-full}  we study how $U$ and $J_{AF}$
change the picture.
We end with our conclusions in Section \ref{concl}.

\section{Truncation of Green's Function Hierarchy} 
\label{formul}
 
In this section we obtain the equations of motion  that determine 
the  Green's function
Eq.(\ref{green-S}) and 
the 
spin susceptibility.
The many--body interactions 
 $H_{exch}$ and $H_U$ introduce an infinite hierarchy 
of coupled equations of motion that involve  higher Green's 
functions of the 
 form 
\begin{equation}
\label{green-A}
\ll A \gg
=-i\theta(t)<[A(t),S_{\bf Q}^{-}(0)]>,
\end{equation}
where $A(t)
= \exp{(i H t)} A
\exp{(-iHt)}$
are many--body Heisenberg operators.
To truncate this hierarchy,
we approximate 
the higher Green's functions by  systematically adding 
correlations among any given number of 
elementary excitations.
To lowest order,
the RPA describes uncorrelated  quasi--particles. 
At the next level, we include
all correlations between any two elementary excitations, 
which determine the inelastic dephasing rate.  

Ref.[\onlinecite{igar}] used a three--body scattering theory 
to 
calculate the 
electron  Green's function of the Hubbard Hamiltonian.
In one dimension, the  results obtained this way were in excellent agreement 
with the exact Bethe ansatz solution.\cite{igar,rucken}    
In Ref.[\onlinecite{perakis}], a similar method was used 
to calculate the density matrix 
that describes the coherent ultrafast nonlinear optical 
dynamics  of the Quantum Hall system.
Refs.[\onlinecite{fes-1,fes-2,fes-3}] calculated the Fermi Edge Singularity 
 in doped semiconductor quantum wells using an analogous approach. 
In this paper, we establish the correspondence 
with a factorization 
scheme of higher Green's functions. 
For simplicity we restrict to zero temperature,  
where Eq.(\ref{green-A}) involves the ground state average value.

In the case of ferromagnetic 
 exchange interaction as in 
the manganites,  
the 
fully polarized 
state
\begin{equation} 
\label{GS}
| F \rangle = 
\prod_{\nu} 
\, c^{\dag}_{\nu \uparrow} |0\rangle \bigotimes | S,S,\cdots \rangle, 
\end{equation} 
is an  {\em exact eigenstate} of the many-body Hamiltonian 
$H$. 
In Eq.(\ref{GS}), 
$|0\rangle$ is the vacuum state 
and $| S,S,\cdots \rangle$
describes local spins with $S_z=S$
on all lattice sites.  From now on, 
the indices $\mu,\nu,\cdots$
denote  states  occupied in $|F\rangle$,  
while $\alpha,\beta,\cdots$ denote
 empty states.
In the 
parameter range of interest here, $|F\rangle$ is the ground state. 
\cite{kapet1,kapet2}
Using the properties  $H |F\rangle =0$ 
(we choose the eigenvalue of $|F\rangle$ as 
the zero of energy)
and 
$\langle F | S^{-}_{{\bf Q}}=0$, both of 
which stem from the fact that 
$|F\rangle$ is the state with maximum spin, 
we obtain from Eq.(\ref{green-A}) 
\begin{equation} 
\ll A\gg=-i \theta(t) <F|Ae^{-iHt}S_{\bf Q}^{-}|F>.
\label{A-amp} 
\end{equation} 
The Green's function
$\ll A\gg$ is then given by the  amplitude of the time--evolved
 state 
$S_{\bf Q}^{-}|F>$.
 The hierarchy of Green's function equations of motion is
equivalent to solving the time--dependent Schr\"{o}dinger equation. 
However, Green's functions
also treat dephasing and relaxation, and can 
describe phenomenologically the effects of 
coupling to degrees of freedom not included 
in the Hamiltonian $H$  by introducing 
phenomenological damping rates.
The coupling of 
$\ll A\gg$
to higher Green's functions
is determined  by the  states $H A |F\rangle$. 
Truncation of the equations of motion 
hierarchy can  be achieved by 
expanding $H A |F\rangle$ in a truncated  basis,    
which gives the exact solution within a subspace of states. 

We start with the equation of motion for the spin Green's function
Eq.(\ref{green-S}), 
obtained 
after straightforward algebra by 
using Eq.(\ref{A-amp}) and  the properties of  
$|F\rangle$:
\begin{eqnarray}\label{eomY}
&&\left(i\partial_{t} - \frac{Jn}{2}
-
\omega^{\rm{AF}}_{\bf Q}
\right)
\ll 
S_{\bf -Q}^{\dagger}\gg \nonumber 
\\
&& =2S\delta(t)-\frac{JS}{\sqrt{N}}\sum_{\nu}\ll c_{\nu\uparrow}^{\dagger}
c_{\bf \nu+Q\downarrow}\gg\nonumber\\
&&+\frac{J}{2N}\sum_{\alpha\nu}\ll 
S_{\bf \alpha-\nu-Q}^{\dagger}c_{\bf \nu\uparrow}^{\dagger}
c_{\bf \alpha\uparrow}\gg,
\end{eqnarray} 
and 
\begin{equation} 
\omega^{\rm{AF}}_{\bf Q}
=2 J_{\rm{AF}}S(\gamma_{\bf Q}-\gamma_{\bf 0})
\end{equation} 
is the spin--wave energy  due to  $H_{\rm{AF}}$.  
The same result can alternatively be  obtained 
by decomposing the Green's 
functions contributing to 
 $\ll [S^+_{{\bf -Q}},H] \gg$ into correlated and uncorrelated 
parts after using the identity 
\begin{eqnarray} 
&& \ll S^n_{{\bf q}} c^\dag_{{\bf k-Q-q} \sigma} 
c_{{\bf k}\sigma^{\prime}}
\gg    \nonumber \\
&&= \langle c^\dag_{{\bf k-Q-q} \sigma} 
c_{{\bf k}\sigma^{\prime}} \rangle 
 \ll  S^n_{{\bf q}}\gg 
 \nonumber\\
&&+
\langle S^n_{{\bf q}} \rangle 
\ll  
c^\dag_{{\bf k-Q-q} \sigma} 
c_{{\bf k}\sigma^{\prime}}\gg \nonumber \\
&&
+ \ll  \Delta S^n_{{\bf q}} 
\Delta [ c^\dag_{{\bf k-Q-q} \sigma} 
c_{{\bf k}\sigma^{\prime}}]
\gg,
\label{G-factor} 
\end{eqnarray}
where $S^n$ are the components 
of the local spin and 
$\Delta A = A - \langle A \rangle $ describes 
the quantum fluctuations of $A$.  
The Green's function $\ll  \Delta S^n 
\Delta [ c^\dag_{\sigma} 
c_{\sigma^{\prime}}]
\gg$
is the correlated part 
of 
$\ll  S^n c^\dag_{\sigma} 
c_{\sigma^{\prime}}
\gg$. 

The Green's function $\ll c_{\nu\uparrow}^{\dagger}
c_{\bf \nu+Q\downarrow}\gg$ describes
the itinerant carrier spin dynamics and 
satisfies the following  equation of motion, 
obtained 
after straightforward algebra by 
using Eq.(\ref{A-amp}) and the properties of 
$|F\rangle$:
\begin{eqnarray}
&&(i\partial_{t}
-\varepsilon_{\bf \nu+Q}+\varepsilon_{\bf \nu}-JS-nU)
\ll c_{\nu\uparrow}^{\dagger}c_{\bf \nu+Q\downarrow}\gg=
\nonumber \\
&&-\frac{J}{2\sqrt{N}}\ll S_{\bf -Q}^{\dagger}\gg-\frac{U}{N}
\sum_{\mu}
\ll c_{\mu\uparrow}^{\dagger}c_{\bf \mu+Q\downarrow}\gg
\nonumber\\
&&-\frac{J}{2\sqrt{N}}\sum_{\bf \alpha}\ll 
S_{\bf \alpha-\nu-Q}^{\dagger}c_{\bf \nu\uparrow}^{\dagger}
c_{\bf \alpha\uparrow}\gg \nonumber \\
&&-\frac{U}{N}
\sum_{\alpha\mu}
\ll c_{\bf \mu\uparrow}^{\dagger}c_{\bf Q+\mu+\nu-\alpha\downarrow}
c_{\bf \nu\uparrow}^{\dagger}c_{\bf \alpha\uparrow}\gg_c.\label{eomX}
\end{eqnarray}
where we defined the correlated part of the four--particle 
Green's function as
\begin{eqnarray} 
&& \ll c^\dag_1 c_2 c^\dag_3 c_4 \gg_c
=\ll c^\dag_1 c_2 c^\dag_3 c_4 \gg
 \nonumber\\
&&- \langle c^\dag_1 c_2 \rangle 
\ll c^\dag_3 c_4 \gg
- \langle c^\dag_3 c_4 \rangle 
\ll  c^\dag_1 c_2 \gg
\nonumber \\ &&
 - \langle c_2  c^\dag_3 \rangle 
\ll  c^\dag_1 c_4 \gg
+\langle c^\dag_1 c_4 \rangle 
\ll  c^\dag_3 c_2 \gg.
\label{Phi-factor}
\end{eqnarray} 
Alternatively, Eq.(\ref{eomX})  can be derived by
using Eq.(\ref{G-factor})
to decompose the Green's function 
$\ll [c_{\nu\uparrow}^{\dagger}c_{\bf \nu+Q\downarrow},H]\gg$.
The first line on the right hand side (rhs)
 of Eqs.(\ref{eomY}) and 
(\ref{eomX}) gives the RPA result, 
which neglects all non--factorizable
(correlated) contributions to Eqs.(\ref{G-factor}) and 
(\ref{Phi-factor})
(Tyablikov approximation \cite{tyablikov}).

Eqs.(\ref{eomY}) and (\ref{eomX}) reduce the calculation 
of the  spin Green's function to that of two higher Green's functions, 
$\ll 
\Delta S^{+} \Delta[c_{\bf \uparrow}^{\dagger}
c_{\bf \uparrow}]\gg$ and 
$\ll c_{\bf \uparrow}^{\dagger}c_{\bf \downarrow}
c_{\bf \uparrow}^{\dagger}c_{\bf \uparrow}\gg_c$.
In a  system 
with a general ground state, 
two additional 
Green's functions, 
$\ll 
\Delta S^{z} \Delta[c_{ \uparrow}^{\dagger}
c_{ \downarrow}]\gg$ 
and $\ll \Delta S^{+} \Delta[c_{\downarrow}^{\dagger}
c_{ \downarrow}]\gg$, also couple and describe 
ground state correlations.  
\cite{unpubl}
However, these vanish here since, 
for the groud state Eq.(\ref{GS}), 
$\Delta S^{z} |F \rangle$=0 and 
$c_{\downarrow}^{\dagger}
c_{ \downarrow}|F\rangle$=0. 
For the same reason,
$\ll c_{\bf \mu\uparrow}^{\dagger}c_{\downarrow}
c_{\bf \nu\uparrow}^{\dagger}c_{\bf \alpha\uparrow}\gg 
= \ll c_{\bf \mu\uparrow}^{\dagger}c_{\downarrow}
c_{\bf \nu\uparrow}^{\dagger}c_{\bf \alpha\uparrow}\gg_c$ 
and 
$\ll 
S^{+}c_{\bf \nu\uparrow}^{\dagger}
c_{\bf \alpha\uparrow}\gg
= \ll 
\Delta S^{+} \Delta[c_{\bf \nu\uparrow}^{\dagger}
c_{\bf \alpha\uparrow}]\gg$. Also, from Eq.(\ref{A-amp}) we see that 
$\ll A \gg=0$ for  any $A$ such that 
$\langle F | A =0$.

The 
Green's function $\ll 
S^{+}c_{\bf \nu\uparrow}^{\dagger}
c_{\bf \alpha\uparrow}\gg$
describes the correlations between a magnon, 
an electron, and a Fermi sea hole (three--body correlations), 
while the Green's function  
$\ll c_{\bf \mu\uparrow}^{\dagger}c_{\bf Q+\mu+\nu-\alpha\downarrow}
c_{\bf \nu\uparrow}^{\dagger}c_{\bf \alpha\uparrow}\gg_c$
describes the correlations between two Fermi sea holes and two electrons
of opposite spin. 
These two Green's functions  are obtained 
from the following equations of motion,
derived from 
Eq.(\ref{A-amp})
after 
using the  properties of 
$|F\rangle$:
\begin{eqnarray}\label{eomG}
&&\left(i\partial_{t}
-\varepsilon_{\bf \alpha}+\varepsilon_{\bf \nu}-\frac{Jn}{2}
-\omega^{\rm{AF}}_{{\bf Q} + \nu - \alpha} \right) 
\ll S_{\bf \alpha-\nu-Q}^{\dagger}
c_{\bf \nu\uparrow}^{\dagger}c_{\bf \alpha\uparrow}\gg \nonumber \\
&&= 
\frac{J}{2N}\ll S_{\bf -Q}^{\dagger}\gg
-
\frac{JS}{\sqrt{N}}\ll c_{\bf \nu\uparrow}^{\dagger}c_{\bf \nu+Q\downarrow}
\gg \nonumber\\
&&+\frac{J}{2N}
\sum_{\bf \beta} \ll S_{\bf \beta-\nu-Q}^{\dagger}c_{\bf \nu\uparrow}^{\dagger}
c_{\bf \beta\uparrow}\gg\nonumber\\
&&
-\frac{J}{2N}\sum_{\bf \mu} \ll S_{\bf \alpha-\mu-Q}^{\dagger}
c_{\bf \mu\uparrow}^{\dagger}
c_{\bf \alpha\uparrow}\gg \nonumber\\
&&-\frac{JS}{\sqrt{N}}\sum_{\bf \mu} 
\ll c_{\bf \mu\uparrow}^{\dagger}c_{\bf Q+\mu+\nu-\alpha\downarrow}
c_{\bf \nu\uparrow}^{\dagger}c_{\bf \alpha\uparrow}\gg_c
\end{eqnarray}
and 
\begin{eqnarray}\label{eomPhi}
&&\left(
i\partial_{t}
-\varepsilon_{\bf Q+\mu+\nu-\alpha}-
\varepsilon_{\bf \alpha}+\varepsilon_{\bf \nu}+
\varepsilon_{\bf \mu}-JS -nU\right) \times \nonumber\\
&&\ll c_{\bf \mu\uparrow}^{\dagger}c_{\bf Q+\mu+\nu-\alpha\downarrow}
c_{\bf \nu\uparrow}^{\dagger}c_{\bf \alpha\uparrow}\gg_c 
\nonumber \\
&&=\frac{J}{2\sqrt{N}}\left[ \ll S_{\bf \alpha-\mu-Q}^{\dagger}
c_{\bf \mu\uparrow}^{\dagger}c_{\bf \alpha\uparrow}\gg
-\ll S_{\bf \alpha-\nu-Q}^{\dagger}
c_{\bf \nu\uparrow}^{\dagger}c_{\bf \alpha\uparrow}\gg\right] \nonumber \\
&&+ \frac{U}{N} \left[
\ll c_{\bf \mu\uparrow}^{\dagger}c_{\bf \mu+Q\downarrow}
\gg_c - \ll c_{\bf \nu\uparrow}^{\dagger}c_{\bf \nu+Q\downarrow}
\gg \right] \nonumber \\
&&- \frac{U}{N} \left[ \sum_{\mu^\prime} 
\ll c_{{\bf \mu^\prime}\uparrow}^{\dagger}
c_{\bf Q+\mu^\prime+\nu-\alpha\downarrow}
c_{\bf \nu\uparrow}^{\dagger}c_{\bf \alpha\uparrow}\gg_c \right. \nonumber \\
&&\left.
+ \sum_{\nu^\prime} 
\ll c_{{\bf \mu}\uparrow}^{\dagger}
c_{\bf Q+\mu+\nu^\prime-\alpha\downarrow}
c_{\bf \nu^\prime\uparrow}^{\dagger}c_{\bf \alpha\uparrow}\gg_c \right.
 \nonumber \\ 
&& \left.- \sum_{\alpha^\prime} 
\ll c_{{\bf \mu}\uparrow}^{\dagger}
c_{\bf Q+\mu+\nu-\alpha^\prime\downarrow}
c_{\bf \nu\uparrow}^{\dagger}c_{\bf \alpha^\prime\uparrow}\gg_c 
\right]. 
\end{eqnarray}
Eqs.(\ref{eomG}) and 
(\ref{eomPhi}) describe vertex corrections to the
carrier--spin interaction. The first line on the rhs of Eq.(\ref{eomG})
gives the Born approximation.
The second and third lines 
describe vertex corrections due to  the multiple scattering 
of the localized spin with the Fermi sea pair 
electron (second line) and hole (third line). 
Neglecting the third line corresponds to  assuming a 
non--interacting (static \cite{fes-1}) Fermi sea,
equivalent to summing only the
electron--magnon  ladder 
diagrams (two--body ladder approximation).\cite{igar} 
By also including the hole multiple scattering processes
(third line), 
we treat exactly all correlations between local spin, 
electron, and hole, a three--body  problem.
The last line on the rhs of Eq.(\ref{eomG}) 
comes from  correlations between 
two electrons and two holes,
described  by Eq.(\ref{eomPhi}). 
We note that $U$ 
introduces new correlations among  all 
four of the above particles, 
 described by the last four lines  on the rhs of 
Eq.(\ref{eomPhi}).  

To obtain the above closed system of  equations, 
we neglected 
the coupling 
to  Green's functions of the form 
$\ll A c^{\dagger}_{\nu \uparrow} 
c_{\mu \uparrow}^{\dagger} c_{\alpha \uparrow}
c_{\beta \uparrow}
\gg$, 
where $A=S^+$ or $c^{\dagger}_{\nu^\prime \uparrow} 
c_{\downarrow}$. These 
neglected Green's functions 
describe 
multi--particle correlations between 
{\em two} Fermi sea pairs
 and 
a local spin or carrier spin--flip excitation, which 
contribute to higher order in 1/S.   
Alternatively, we can arrive at the same result by 
decomposing the Green's functions
$\ll S^+ c^{\dagger}_{ \uparrow} 
c_{ \uparrow}^{\dagger} c_{\uparrow}
c_{ \uparrow}
\gg$ and 
$\ll c^{\dagger}_{\uparrow} 
c_{\downarrow}  c^{\dagger}_{ \uparrow} 
c_{ \uparrow}^{\dagger} c_{\uparrow}
c_{ \uparrow}
\gg$ 
into uncorrelated and  correlated parts, 
by separating out all possible 
factorizable 
contributions 
 similar to 
Ref.[\onlinecite{corr-exp}],  
and neglecting the fully correlated contributions that  
describe correlations among three excitations.  
This correlation expansion 
neglects the contribution of 
 states with two or more 
Fermi sea pair excitation and 
corresponds 
to an exact 
calculation of the Green's functions within the sub--space of states 
with up to one Fermi sea pair.
As discussed e.g. in Refs.[\onlinecite{igar,rucken}]
and implied by Eq.(\ref{A-amp}),  
in the limit $\gamma$,$\Gamma$$\rightarrow$0 
the exact calculation  of the Green's function
within a given subspace 
it equivalent to the variational 
calculation of the spin--wave energy
using a variational wavefunction 
that is a linear combination of the  states
that span the subspace
(obtained in 
Refs.[\onlinecite{kapet1,kapet2}] for the problem at hand).
One could include multipair correlations, 
e.g., by extending the approach of 
Refs. [\onlinecite{linden},\onlinecite{fes-2},
\onlinecite{fes-3}].

\section{Spin Self Energy} 
\label{spin-self} 
 
The spin self energy can be calculated by solving the 
 equations of motion derived  
above  by Fourier transformation. 
Eq.(\ref{eomY}) gives 
\begin{eqnarray}\label{eomY-F}
\ll 
S_{\bf -Q}^{\dagger}\gg_{\omega}=
\frac{2S}{\omega 
-\omega^{\rm{AF}}_{\bf Q}
- \Sigma(\omega,{\bf Q})
},
\end{eqnarray} 
where $\Sigma(\omega,{\bf Q})$
is the self energy.
Defining for convenience 
\begin{eqnarray} 
&&X_{\nu}(\omega,{\bf Q}) = \frac{\ll c_{\nu\uparrow}^{\dagger}
c_{\bf\nu+Q\downarrow}\gg_\omega}{
\ll S_{\bf -Q}^{+}\gg_\omega}, 
\label{def-X}\\
&& G_{\alpha\nu}(\omega,{\bf Q})= 
\frac{\ll S_{\bf \alpha-\nu-Q}^{\dagger}c_{\bf \nu\uparrow}^{\dagger}
c_{\bf \alpha\uparrow}\gg_\omega}{\ll S_{\bf -Q}^{+}
\gg_\omega}, 
\label{def-G} \\
&&\Phi^\alpha_{\mu\nu}(\omega,{\bf Q})= \frac{\ll c_{\mu\uparrow}^{\dagger}
c_{\bf Q+\mu+\nu-\alpha\downarrow}
c_{\nu\uparrow}^{\dagger}c_{\alpha\uparrow}\gg_\omega}{
\ll S_{\bf -Q}^{+}\gg_\omega}\label{def-Phi} 
 \end{eqnarray} 
and  substituting
into Eqs.(\ref{eomY}), (\ref{eomX}), (\ref{eomG}), and (\ref{eomPhi}) we 
express the self energy 
in the form 
\begin{eqnarray}
\label{self}
\Sigma(\omega,{\bf Q})=  \frac{Jn}{2}
-\frac{\Delta}{\sqrt{N}}
\sum_{\nu}X_{\nu}
+\frac{J}{2N}\sum_{\alpha\nu}G_{\alpha\nu},
\end{eqnarray}
where $\Delta = J S$ is the magnetic energy.
We calculate $\Sigma$
non--perturbatively in the interactions and $1/S$ 
by solving the following coupled equations for $X$, $G$, and $\Phi$: 
\begin{eqnarray}
&&\left(nU + \Delta 
+\varepsilon_{\bf\nu+Q} -\varepsilon_{\nu}-
\omega \right)
X_{\nu}
- \frac{U}{N} \sum_{\mu}X_{\mu}
\nonumber\\
&&
=\frac{J}{2 \sqrt{N}} \left(1 + 
\sum_{\alpha}G_{\alpha\nu}\right) 
+
\frac{U}{N}\sum_{\alpha\mu}\Phi_{\mu\nu}^{\alpha}, \label{X-F}
\end{eqnarray}
\begin{eqnarray} 
&&\left(\omega+\varepsilon_{\nu}-\varepsilon_{\alpha}-
\frac{Jn}{2}
-\omega^{AF}_{\bf Q -\alpha+\nu} +i\gamma
\right)G_{\alpha\nu}\nonumber\\
&&=\frac{J}{2N}
\left( 1 + \sum_{\beta}
G_{\beta\nu} -
\sum_{\mu} G_{\alpha\mu}\right) \nonumber \\ 
&& - \frac{\Delta}{\sqrt{N}} \left(
X_{\nu} + \sum_{\mu}\Phi_{\mu\nu}^\alpha \right), 
\label{G-F}
\end{eqnarray}
where $\gamma \rightarrow 0$,  
and 
\begin{eqnarray} 
&&\left(nU + \Delta
+\varepsilon_{\alpha}+\varepsilon_{\bf Q+\mu+\nu-\alpha}
-\varepsilon_{\mu}-\varepsilon_{\nu}
-\omega \right)
 \Phi_{\mu\nu}^\alpha \nonumber \\
&& = \frac{J}{2 \sqrt{N}}
\left( G_{\alpha\nu}-G_{\alpha\mu}\right)
+ \frac{U}{N}
\left(X_{\nu}-X_{\mu}\right)\nonumber\\
&&+\frac{U}{N} \left( \sum_{\mu'}\Phi_{\mu'\nu}^\alpha+
\sum_{\nu'}\Phi_{\mu\nu'}^\alpha 
-\sum_{\beta}\Phi_{\mu\nu}^{\beta}
\right) \label{Phi-F}.
\end{eqnarray}

First we consider the 
RPA self energy,  obtained by setting 
 $G=\Phi=0$ in the above equations. 
We can then solve Eq.(\ref{X-F}) analytically 
after noting that its solution has the form 
 \begin{eqnarray}
\label{X-RPA1}
X^{\rm{RPA}}_{\nu}(\omega,{\bf Q})= \frac{ \chi^{\rm{RPA}}(\omega,{\bf Q})}{
nU + \Delta 
+\varepsilon_{\bf\nu+Q} -\varepsilon_{\nu}-
\omega}.
\end{eqnarray} 
Substituting the above expression 
into Eq.(\ref{X-F}) we obtain 
\begin{eqnarray} 
\label{X-RPA2}
\chi^{\rm{RPA}}(\omega,{\bf Q})
= \frac{J}{2 \sqrt{N}} 
\frac{1}{1 - \frac{U}{N} \sum_{\mu} 
\frac{1}{
nU + \Delta + \varepsilon_{\bf\mu+Q}-\varepsilon_{\mu}- \omega}}.
\end{eqnarray} 
which gives after some straightforward algebra 
 \begin{eqnarray}
\label{X-RPA}
X^{\rm{RPA}}_{\nu}(\omega,{\bf Q})= 
\frac{J}{2 \sqrt{N}}
\frac{1}{\Delta + 
U_{{\bf \nu Q}}
+ \varepsilon_{\bf\nu+Q} -\varepsilon_{\nu} 
-
\omega}, 
\end{eqnarray}
where we introduced the Coulomb--induced energy 
\begin{eqnarray} 
U_{{\bf \nu Q}} = 
 \frac{U}{N} \sum_{\mu} 
\frac{\varepsilon_{\bf \mu +Q} -\varepsilon_{\mu}-
\varepsilon_{\bf\nu+Q} +\varepsilon_{\nu}}{
n U + \Delta - i \Gamma + \varepsilon_{\bf \mu +Q} -\varepsilon_{\mu}-
\omega}.\label{U-en} 
\end{eqnarray} 
Substituting Eq.(\ref{X-RPA}) 
into 
Eq.(\ref{self}) after setting $G=0$ we 
obtain the RPA self energy
\begin{equation} 
\label{Sigma-RPA}
\Sigma^{\rm{RPA}}(\omega,{\bf Q})=  \frac{J}{2N}
\sum_{\nu} 
\frac{
U_{{\bf \nu Q}}
+ \varepsilon_{\bf\nu+Q} -\varepsilon_{\nu} 
-
\omega}{\Delta - i \Gamma 
+ U_{{\bf \nu Q}}
+ \varepsilon_{\bf\nu+Q} -\varepsilon_{\nu} 
-
\omega}. 
\end{equation} 
Following Ref.[\onlinecite{heinr}], 
in the above equations we added 
a phenomenological 
relaxation rate 
$\Gamma$
that describes 
the itinerant 
spin lifetime due to interactions not included 
in our Hamiltonian $H$. 
This result can alternatively  
be obtained by substituting $\Delta$ by $\Delta - i \Gamma$, 
as derived with the Lindblad semigroup method in Ref.[\onlinecite{chovan}].
In the intrinsic system described by the Hamiltonian $H$,  
$\Gamma$$\rightarrow$0.

We now turn to the 
self--energy due to the correlations. 
By formally solving Eq.(\ref{X-F}) for $X_{\nu}$ and substituting 
into Eq.(\ref{self}), 
we separate the  self--energy 
into RPA  and 
correlated contributions, 
$\Sigma(\omega,{\bf Q})= \Sigma^{\rm{RPA}}(\omega,{\bf Q}) +
\Sigma^{\rm{corr}}(\omega,{\bf Q})$.  
After some algebra we obtain that
$\Sigma^{\rm{corr}}(\omega,{\bf Q})=
\Sigma^{\rm{corr}}_J(\omega,{\bf Q})+ 
\Sigma^{\rm{corr}}_U(\omega,{\bf Q})$. 
 \begin{eqnarray} 
\label{self-J} 
\Sigma^{\rm{corr}}_J=
\frac{J}{2N} \sum_{\alpha \nu} G_{\alpha \nu} 
\frac{
U_{{\bf \nu Q}}
+ \varepsilon_{\bf\nu+Q} -\varepsilon_{\nu} 
-
\omega}{\Delta - i \Gamma 
+ U_{{\bf \nu Q}}
+ \varepsilon_{\bf\nu+Q} -\varepsilon_{\nu} 
-
\omega}
\end{eqnarray}
is the contribution of the Fermi sea--magnon correlations 
due to $J$, 
described by the Green's function $G$, Eq.(\ref{G-F}).  
 \begin{eqnarray} 
\label{self-U} 
\Sigma^{\rm{corr}}_{U}=
-\frac{U}{N^{3/2}} \sum_{\alpha \nu \mu} \Phi^{\alpha}_{\mu \nu} 
\frac{\Delta-i \Gamma}{
\Delta - i \Gamma 
+ U_{{\bf \nu Q}}
+ \varepsilon_{\bf\nu+Q} -\varepsilon_{\nu} 
-  
\omega}
\end{eqnarray} 
is the contribution
of  the Fermi sea pair--carrier spin--flip four--particle correlations 
described by $\Phi$ and discussed in Ref.[\onlinecite{kapet2}]. 
 This latter contribution vanishes for  $U$=0.  

\section{Numerical results}
\label{num}

\begin{figure}[t]
\vspace{0.35 in}
\centerline{
\hbox{\psfig{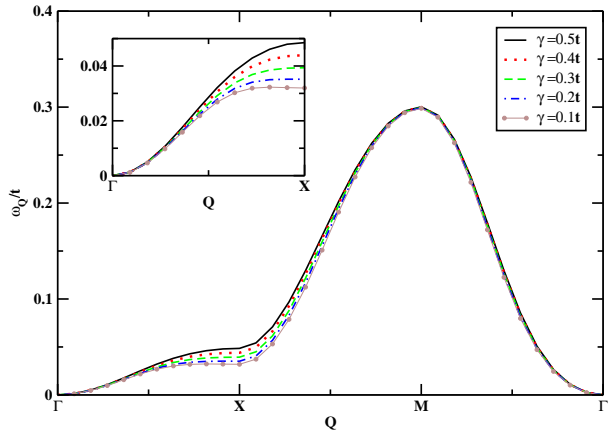}}}
\caption{(Color online) Spin--wave dispersion for different values of 
$\gamma/t$ for $J=8t, n=0.6, U=J_{AF}=\Gamma=0$.  
Inset: Dispersion along the $\Gamma\to X$ direction.}\label{Fig1} 
\end{figure} 

In this section we present our numerical results for the 
spin--wave dispersion, 
which we obtain by 
solving self--consistently the equation 
\begin{eqnarray} 
\omega_{{\bf Q}}= \omega^{\rm{AF}}_{{\bf Q}} 
+ \rm{Re}\Sigma (\omega_{{\bf Q}},{\bf Q}),
\label{omega-def}
\end{eqnarray} 
and the spin--wave dephasing rate
$\Gamma_{{\bf Q}}$, 
determined by 
$\rm{Im}\Sigma (\omega_{{\bf Q}},{\bf Q})$.
We focus on  the inelastic contribution
to the dephasing rate,  
due to 
the scattering between spin and charge excitations: 
$\Gamma_{{\bf Q}}=\Gamma^J_{{\bf Q}}+\Gamma^U_{{\bf Q}}$, 
where 
\begin{eqnarray}
\Gamma^J_{{\bf Q}}
= -\rm{Im} 
\Sigma^{\rm{corr}}_J(\omega_{{\bf Q}},{\bf Q})
\end{eqnarray} 
 is the contribution 
due to the 
magnon--Fermi sea pair correlations described by $G$, 
Eq.(\ref{self-J}), and 
\begin{eqnarray}
\Gamma^U_{{\bf Q}}
= -\rm{Im} 
\Sigma^{\rm{corr}}_U(\omega_{{\bf Q}},{\bf Q})
\end{eqnarray} 
 is  the contribution  
due to the carrier spin flip--Fermi sea pair 
correlations 
described by $\Phi$,  Eq.(\ref{Phi-F}).
An additional elastic contribution to the spin--wave lifetime 
can come from the imaginary part of the RPA self energy, 
which however  vanishes in the limit 
$\Gamma \rightarrow 0$ due to the finite carrier spin--flip excitation energy.
Within the RPA, spin--wave dephasing can only arise from the interplay between
the magnetic  exchange interaction 
and  carrier spin dephasing via external couplings.\cite{mitchell,heinr}

Below we study the momentum 
dependence of $\Gamma_{{\bf Q}}$ 
along different directions in the 
Brillouin zone.
We focus, in particular, on the directions 
$\Gamma$--$X$, $\Gamma$--$M$, 
and $X$--$M$, 
where 
$\Gamma =(0,0), X =(\pi,0)$, and $M =(\pi,\pi)$.
Our calculations were performed 
on a 20$\times$20 square  lattice, 
which as shown in Refs.[\onlinecite{kapet1}] and [\onlinecite{kapet2}] 
gives good convergence  to the thermodynamic limit.
Any small size effects are washed out when the relaxation rate 
$\gamma$ in Eq.(\ref{G-F})  
exceeds the energy spacing.

\subsection{Minimal double--exchange  model}
\label{num-de}

First we consider the simple double exchange Hamiltonian 
and set $U=J_{AF}=\Gamma=0$. 
Fig.\ref{Fig1} shows 
the spin--wave dispersion $\omega_{{\bf Q}}$
as function of the damping rate $\gamma$ 
in Eq.(\ref{G-F}).  
With decreasing $\gamma$, 
our results converge to  the spin--wave dispersion obtained 
variationally in Ref.[\onlinecite{kapet1}] and the 
$\gamma \rightarrow 0$ limit. 
Similar to the experiment,\cite{ye}  our calculation 
can then be fitted 
to the dispersion of  the Heisenberg Hamiltonian 
with first-- and fourth--nearest--neighbor 
spin interactions $J_1$ and $J_4$.\cite{kapet2}
For intermediate concentrations, our calculation gives a 
pronounced spin--wave softening at
the X point, described by $J_4$,   
 as compared to both the RPA 
and to the fit to the nearest neighbor Heisenberg 
dispersion. 
The effects of a finite  $\gamma$
are most pronounced 
along the direction $\Gamma\to X$: 
with increasing $\gamma$, 
the time evolution described by the 
Green's function $G$, 
which determines  the spin--wave softening, 
is suppressed and thus 
the dispersion starts to approach 
the RPA ($G=0$) result.
From now on 
we fix $\gamma =0.2t$, which as can be seen 
in  
Fig.\ref{Fig1} is close to the 
$\gamma \rightarrow 0$ limit.

\begin{figure}[t]
\vspace{0.36 in}
\centerline{
\hbox{\psfig{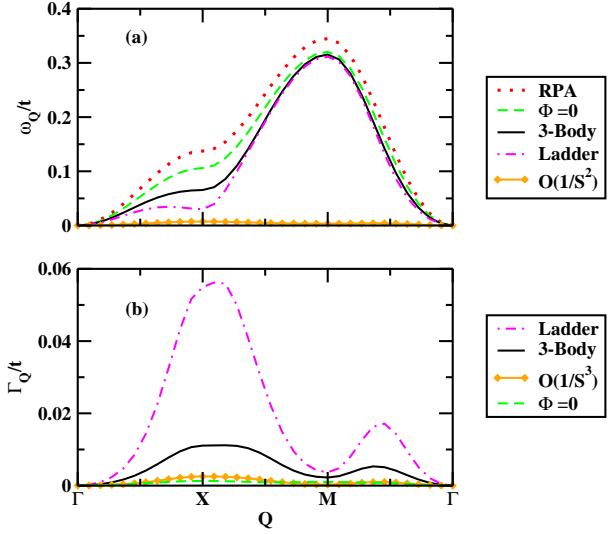}}}
\caption{ (Color online)
Comparison of the different 
approximations for treating the correlations  for $U$=0. 
(a) Spin--wave dispersion, (b) Inelastic spin--wave dephasing 
 rate.
$n =0.6, J =8t, \gamma =0.2t, U=J_{\rm{AF}}=\Gamma=0$.
}\label{Fig2} 
\end{figure}

 Fig.\ref{Fig2}
demonstrates the important role  of  correlations 
due to spin--charge interactions 
on both the spin--wave dispersion  and dephasing rate.  
The spin--wave energies and lifetimes 
differ markedly depending on the approximation used to 
treat the  correlations.
The latter determine the  differences from 
the RPA,
which describes non--interacting spin--waves ($G=\Phi$=0). 
The RPA   
 grossly underestimates the  softening and 
does not give any spin damping in the limit 
$\Gamma \rightarrow 0$.

By neglecting 
the Green's function $\Phi$,  
we obtain spin--wave  energies closer to the RPA (see  Fig.\ref{Fig2}(a)).
As seen in  Fig.\ref{Fig2}(b), this
approximation, which only treats the scattering 
of local spins
with the Fermi sea,  gives a very small damping rate.  
On the other hand, the non--variational $O(1/S^2)$ 
approximation discussed in Refs.[\onlinecite{golosov,chubukov}],  which  
treats magnon--Fermi sea 
scattering  within the Born approximation
and Fermi's golden rule,
strongly overestimates the  softening, while at the same time 
 predicting only a  small damping rate (see  Fig.\ref{Fig2}).

By adding to the 
$O(1/S^2)$ result 
the effects of the multiple 
scattering of the magnon 
with the Fermi sea pair electron,
while still neglecting the magnon--Fermi sea hole  
interactions, 
 we obtain 
a non--variational two--body approximation of the vertex 
corrections equivalent to 
summing the magnon--electron ladder diagrams.\cite{igar}
As can be seen in Fig.\ref{Fig2}, this ladder approximation gives very 
large softening {\em and} damping, 
much larger 
than the predictions of the full calculation
(which is variational in the limit 
$\gamma, \Gamma$$\rightarrow$0 considered here).  
The latter 
treats, in addition to the magnon--electron interactions, 
the 
multiple scattering of the Fermi sea pair hole with the magnon.  
The large differences between the ladder and full calculation 
results demonstrate the importance of three--body correlations 
between magnon, electron, and hole 
in the parameter regime of interest in the manganites. 
We conclude based on Fig.\ref{Fig2} that 
{\em all} correlations between spin, electron, and hole 
 must be treated on an equal basis.
The variational nature of our full calculation 
of the spin--wave energies in the limit 
$\gamma, \Gamma$$\rightarrow$0 
has the advantage 
of  providing  a rigorous limit of the magntitude of the softening,
 unlike for the ladder or 1/S expansion results. 

\begin{figure}[t]
\vspace{0.36 in}
\centerline{
\hbox{\psfig{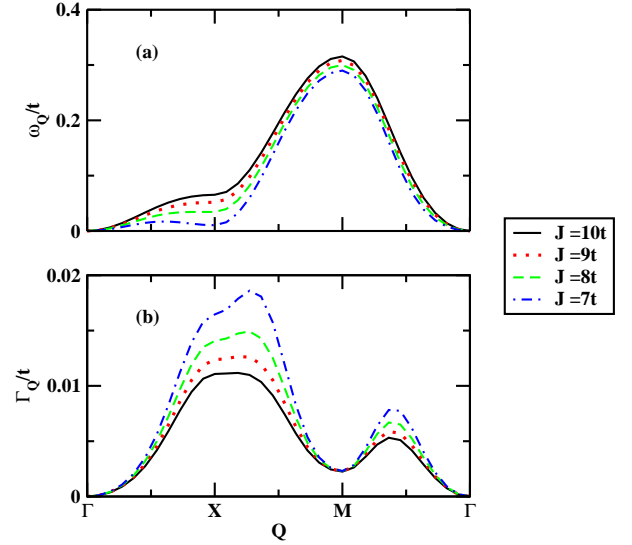}}}
\caption{ (Color online)
Dependence of the spin dynamics 
on the interaction strength $J$. 
 Spin--wave dispersion (a) and dephasing rate (b).
$n=0.6,\gamma=0.2t,\Gamma=U=J_{AF}=0$.}\label{Fig3}
\end{figure}

Fig.\ref{Fig3} 
shows the behavior of $\omega_{{\bf Q}}$ and $\Gamma_{{\bf Q}}$ 
for different interactions $J$. 
With increasing interaction strength,  
the spin--wave energies increase and the ferromagnetic phase becomes 
more stable.  
This hardening with $J$ is accompanied by a corresponding 
increase in the spin--wave lifetime.
The above changes are stronger along the $\Gamma$--$X$ 
direction, where the non--Heisenberg behavior and softening 
are  pronounced,   and depend nonlinearly on 
$J$/$t$. 
The overall 
momentum dependence, however,  remains the same for all $J$.

\begin{figure*}[t]
\vspace{0.38 in}
\centerline{
\hbox{\psfig{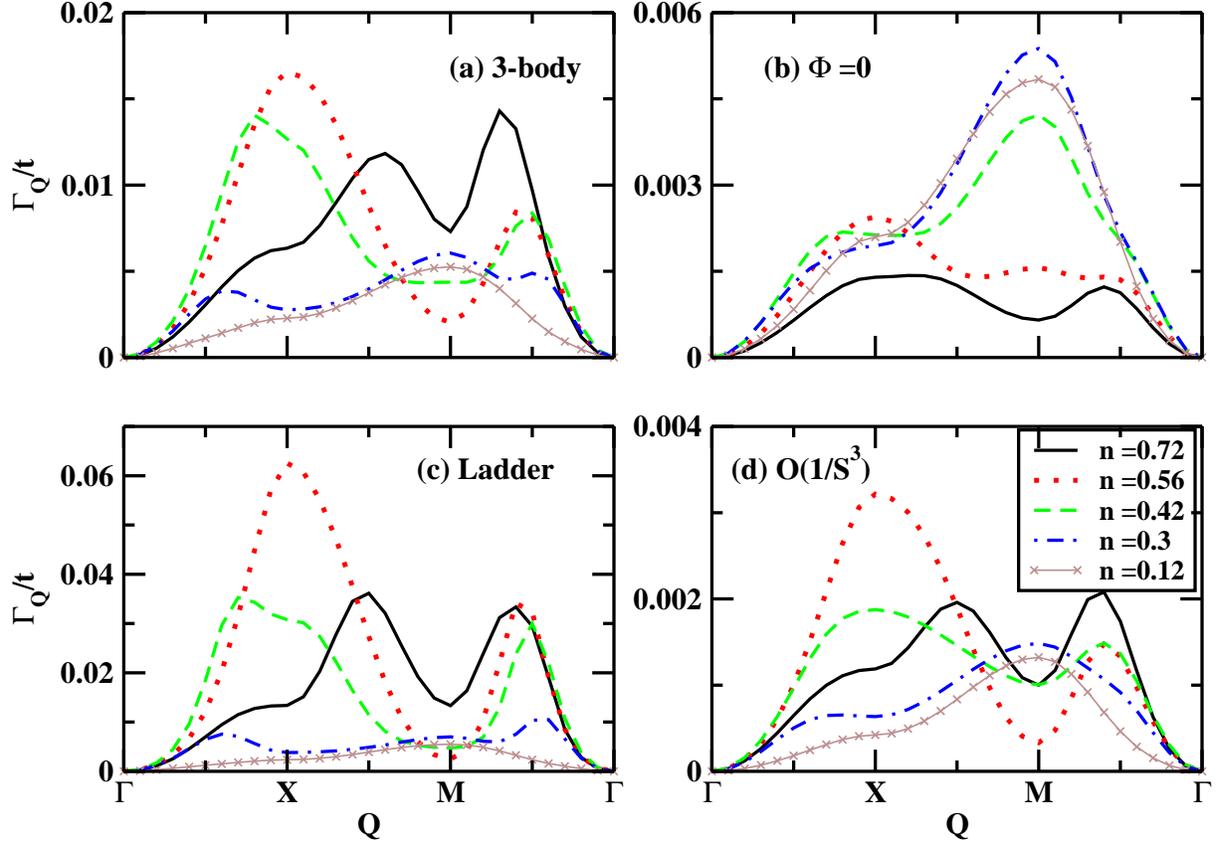}}}
\caption{
(Color online)
Dependence of the spin--wave dephasing rate 
on the carrier concentration $n$ for $U=0$:   
comparison between the different approximations 
of the correlations 
$J =8t, \gamma=0.2t, \Gamma=U=J_{\rm{AF}}=0$.\label{Fig4}}
\end{figure*}

To interpret the  above results, 
we turn to the Green's function 
equations of motion and note that, 
for $U$=0, 
$\Sigma^U$=0. 
After solving 
Eqs.(\ref{X-F}) and (\ref{Phi-F})
for $X$ and $\Phi$  and substituting 
into Eq.(\ref{G-F}) 
we obtain that 
\begin{eqnarray}
\label{G-F-1}
&&\Omega_{\bf \alpha\nu}
G_{\alpha\nu} \nonumber =\\
&&\frac{J}{2N} 
\frac{\varepsilon_{\bf \nu+Q}-\varepsilon_{\nu}-\omega}
{\Delta - i \Gamma +\varepsilon_{\bf \nu+Q} - \varepsilon_{\nu}
- \omega}
\left(1+\sum_{\beta}G_{\beta\nu}\right) \nonumber\\
&&-\frac{J}{2N}\sum_{\mu} G_{\alpha\mu} \times \nonumber \\
&&\frac{\varepsilon_{\bf Q+\mu+\nu-\alpha}
-\varepsilon_{\mu} + \varepsilon_{\alpha}-\varepsilon_{\nu} - \omega}{
\Delta - i \Gamma +
\varepsilon_{\bf Q+\mu+\nu-\alpha}
-\varepsilon_{\mu} + \varepsilon_{\alpha}-\varepsilon_{\nu} - \omega},
\end{eqnarray}
where
\begin{eqnarray}\label{Dav}
&&\Omega_{\bf \alpha\nu}=\omega +i\gamma 
-\left(\varepsilon_{\alpha}-\varepsilon_{\nu}\right) 
\nonumber 
\\
&&-\frac{J}{2N}
\sum_{\mu}
\frac{\varepsilon_{\bf Q+\mu+\nu-\alpha}
-\varepsilon_{\mu} 
+\varepsilon_{\alpha}-
\varepsilon_{\nu}-\omega}{\Delta - i \Gamma+ \varepsilon_{\bf Q+\mu+\nu-\alpha}
-\varepsilon_{\mu} 
+\varepsilon_{\alpha}-
\varepsilon_{\nu}-\omega}.
\end{eqnarray}
and $\gamma, \Gamma$ $\rightarrow$0. 

Spin--wave dephasing results from 
the scattering of  the magnon of momentum ${\bf Q}$ 
to momentum ${\bf Q}+{\bf \nu} - {\bf \alpha}$
while an electron is excited from the state ${\bf \nu}$ 
inside the Fermi sea to the empty state ${\bf \alpha}$. 
In the limit $\gamma \rightarrow 0$,  
this magnon--Fermi sea scattering process must satisfy
the energy conservation condition 
 $\Omega_{\bf \alpha\nu}=0$,  
i.e. 
the initial magnon energy, $\omega=\omega_{{\bf Q}}$,  
must equal the final state energy that 
includes 
the Fermi sea pair energy and the 
${\bf Q}+{\bf \nu} - {\bf \alpha}$ magnon energy.
The final state magnon energy, 
given by the last term in Eq.(\ref{Dav}), 
comes from the coupling of $G$ to $\Phi$.
For $\Phi$=0, 
this spin--wave energy is replaced 
by the  local spin 
excitation energy $Jn/2$. 
The scattering of small energy 
Fermi sea pair excitations from right below to right above 
the Fermi surface  dominates the spin--wave lifetime. 
The density of states and characteristic momenta of such pair excitations 
depends on the shape of the Fermi surface and therefore on the 
carrier concentration. 

To derive the
O($1/S^3)$ 
dephasing rate,   \cite{golosov,chubukov} 
we neglect all rescattering terms 
 on the rhs of Eq.(\ref{G-F-1}) ($\propto G$).
Substituting the expression for $G$ obtained this way 
into Eq.(\ref{self-J}), we  obtain the Born approximation self energy  
 \begin{eqnarray} 
\label{self-J-B} 
\Sigma^{\rm{B}}_J=
\frac{J^2}{4N^2} \sum_{\alpha \nu} 
\left(\frac{
 \varepsilon_{\bf\nu+Q} -\varepsilon_{\nu} 
-
\omega}{\Delta            
+ \varepsilon_{\bf\nu+Q} -\varepsilon_{\nu} 
-
\omega}\right)^2 \frac{1}{\Omega_{\alpha \nu}}. 
\end{eqnarray}
Expanding 
 in terms of  1/S while keeping 
$\Delta=JS$ fixed and 
substituting $\omega=\omega^{(1)}_{{\bf Q}}$, where 
$\omega^{(1)}_{{\bf Q}}$ denotes  
the $O(1/S)$ spin--wave energy, 
 we 
obtain the lowest order 
contribution to 
the self--energy imaginary part:
\begin{equation}\label{damping-1/S}
\rm{Im} \Sigma_J^{\rm{B}}(\omega_{{\bf Q}},{\bf Q}) 
\approx \frac{\Delta^{2}}{4N^{2}S^{2}}\sum_{\alpha\nu}
\left(\frac{\varepsilon_{\nu}-\varepsilon_{\nu+\bf Q}}{
\varepsilon_{\nu}-\varepsilon_{\nu+\bf Q}-\Delta}
\right)^2{\rm Im} \frac{1}{\Omega_{\alpha\nu}}, 
\end{equation}
where 
\begin{eqnarray}
\Omega_{\bf \alpha\nu} = \omega^{(1)}_{{\bf Q}}
-\left(\varepsilon_{\alpha}-\varepsilon_{\nu}
+ \omega^{(1)}_{{\bf Q}-\alpha+\nu} \right) 
- i \gamma.
\end{eqnarray}
The above result corresponds to the Fermi's golden rule
description of the magnon lifetime.
Its large difference from our full calculation, 
demonstrated by Fig.\ref{Fig2},
is due  to the magnon--electron and magnon--hole
multiple interactions (vertex corrections), 
described by the terms proportional to $G$ 
on the rhs of Eq.(\ref{G-F-1}). 
The comparison between the different approximations 
shows that,  
in the parameter 
regime of interest in the manganites, 
the vertex corrections due to three--body correlations 
renormalize significantly the magnon--carrier scattering.

We finally turn 
to the dependence of the  spin 
relaxation  on the carrier concentration.
Fig.\ref{Fig4}(a) demonstrates 
a strong $n$--dependence 
of $\Gamma_{{\bf Q}}$, which 
correlates 
with an 
analogous dependence of $\omega_{{\bf Q}}$ 
and the spin--wave softening discussed in Ref.[\onlinecite{kapet2}]. 
As $n$ decreases 
and the softening (non--Heisenberg behavior)  disappears, 
the spin--wave lifetime increases while its momentum dependence 
changes. 
For intermediate $n$, 
$\Gamma_{{\bf Q}}$ 
displays 
two sharp peaks and a dip 
as function of momentum.
For small $n$, the overall $\Gamma_{{\bf Q}}$  decreases 
and the positions of its maxima and 
minima change.
  
To see this concentration dependence in more detail, we note that,  
for $n=0.72$, 
the spin--wave damping is maximized 
for ${\bf Q}\sim(\pi,\pi/2)$, between $X$ and $M$,  
and  ${\bf Q}\simeq(\pi,\pi/2)$, between $M$ and $\Gamma$,  
while it is minimized close 
 to the $M$ point.
As $n$ decreases to intermediate values, 
the first of the above maxima  approaches
the $X$--point
 while the second maximum  shifts closer to
${\bf Q}=(\pi,\pi/2)$. 
 For smaller $n$, 
the dip 
close to the $M$ point 
turns into a maximum. 
As a result 
of this $n$--dependence, 
$\Gamma_{{\bf Q}} /\omega_{{\bf Q}}$
becomes  quite large in the direction $\Gamma \to X$
for intermediate $n$,
which implies that these spin--wave quasi--particles
interact strongly with the Fermi sea.  
For small $n$, 
$\Gamma_{{\bf Q}}/\omega_{{\bf Q}}$ decreases again. 
The  changes in the 
momentum dependence with $n$ 
are related  to the 
changes in the shape and position of the  Fermi surface, 
which is located close to the Brillouin zone boundary
for  the  higher $n$
but moves towards the center of the Brillouin zone 
as $n$ decreases. 
As a result, the phase space available 
for magnon--carrier scattering changes 
drastically with $n$.

Fig.\ref{Fig4} also compares 
the concentration dependence of 
$\Gamma_{{\bf Q}}$ predicted by the  different 
approximations of the spin--charge interactions. 
By comparing the full three--body calculation 
with the O($1/S^3$) Fermi's golden rule result,
it is clear that the spin--wave damping is 
grossly underestimated by the perturbative 
1/S expansion for all concentrations. 
 Figs.\ref{Fig4}(b), obtained by setting $\Phi$=0,  
fails completely  to capture the correct 
concentration dependence (compare Figs.\ref{Fig4}(a) and 
\ref{Fig4}(b)). It also predicts very small
dephasing rates  for all $n$. 
The above approximation 
neglects the 
interactions between Fermi sea pair and  carrier spin--flip 
excitations.  We therefore conclude that 
such carrier--carrier interactions 
strongly affect 
the magnetization relaxation.
Finally, the comparison of 
 Figs.\ref{Fig4}(a) and (c)  shows that the two--body ladder 
approximation grossly overestimates the spin--wave damping
for intermediate or high $n$, 
while the discrepancies from the full three--body calculation 
decrease for small $n$. 
We conclude based on 
Fig.\ref{Fig4} that the collective spin 
relaxation predicted by the minimal double--exchange model 
can be controlled by tuning the carrier 
concentration $n$, by doping or with external probes. Such 
tuning is heavily influenced by  
 the correlations, which must be treated 
accurately in order 
to capture even the correct order of magnitude and momentum 
dependence of the spin dephasing rate for all concentrations.

\begin{figure}[t]
\vspace{0.38 in}
\centerline{
\hbox{\psfig{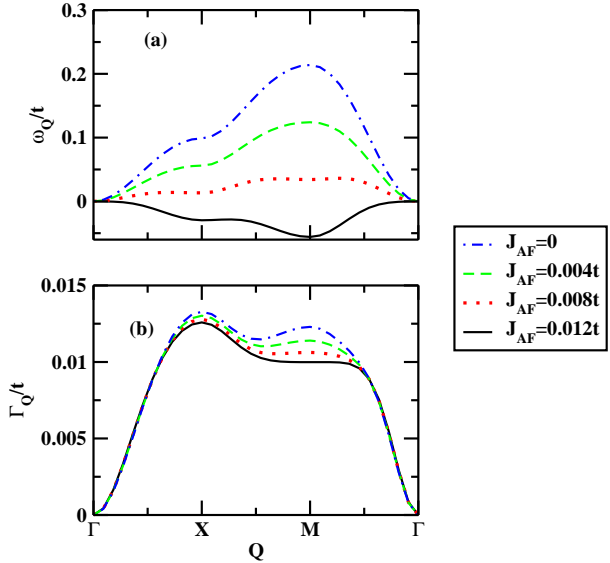}}}
\caption{(Color online) Role of the super--exchange interaction
$J_{\rm{AF}}$. (a) Spin--wave dispersion and (b) Spin--wave damping rate.
$J=2t, U=10t, n=0.6, \Gamma=0.5t,\gamma =0.2 t $}\label{Fig5}
\end{figure}

\subsection{The role of the Coulomb Repulsion}
\label{num-full}

In this section we study how 
 the on--site Coulomb repulsion, $U$, 
and  direct superexchange interaction, $J_{AF}$, 
affect  the spin--wave energies and lifetimes.
Fig.\ref{Fig5} compares 
the results obtained 
for different values of $J_{\rm{AF}}/t$ within the 
range  $0\le J_{AF}\le 0.012t$ relevant to the manganites.\cite{dagotto} 
$J_{\rm{AF}}$ leads to an overall softening of the spin--wave energies. These
eventually turn negative, 
implying instability of the fully polarized 
ferromagnetic phase.
However, $J_{AF}$ preserves 
the nearest neighbor Heisenberg  model 
momentum dependence, 
unlike for the softening observed 
in the experiment.\cite{ye} 
Also, it 
does not affect the spin damping in a significant way. 
We take 
$J_{\rm{AF}}=0$ from now on.

\begin{figure}[t]
\vspace{0.38 in}
\centerline{
\hbox{\psfig{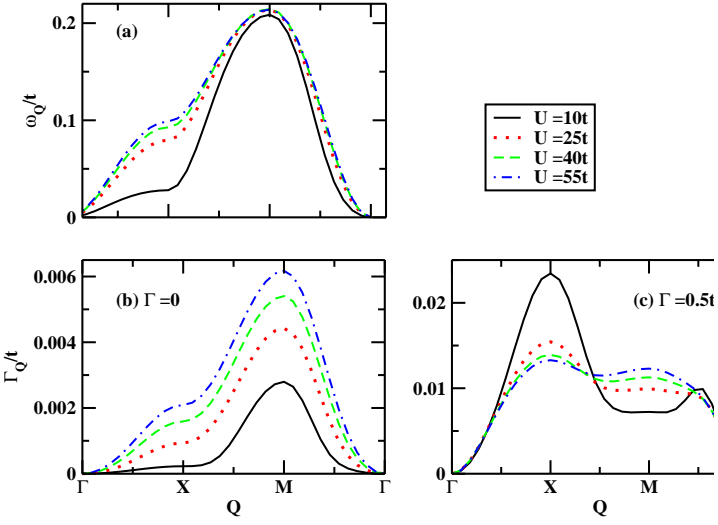}}}
\caption{(Color online) 
Effects of the on--site Coulomb  repulsion $U$. 
(a) Spin--wave dispersion, (b) Spin--wave damping rate
for $\Gamma=0$, and (c) (b) Spin--wave damping rate for $\Gamma$=0.5$t$. 
$n =0.6, J =2t, \gamma =0.2t, J_{\rm{AF}} =0 $.}\label{Fig6}
\end{figure}

As demonstrated by  
Fig.\ref{Fig6}, the effects of the on--site Coulomb (Hubbard) 
repulsion $U$ are more significant. 
Fig.\ref{Fig6}(a)
shows the dependence of the spin--wave dispersion on $U$. 
With increasing $U$, the spin--wave 
softening and deviations from the Heisenberg model dispersion 
diminish as the ferromagnetic phase becomes more stable. 
As discussed in Ref.[\onlinecite{kapet2}], 
our calculated dispersion can be fitted to the 
Heisenberg model dispersion with first-- and fourth--nearest--neighbor
interactions, similar to the experiment. \cite{ye}
Fig.\ref{Fig6}(b)  demonstrates 
 qualitative changes in the overall momentum dependence of the
spin--wave damping as compared to the minimal double exchange 
model. 
In particular, if the carrier spin is conserved ($\Gamma$=0), 
 $\Gamma_{{\bf Q}}$
is maximum at the $M$ point, 
while the damping at the $X$ point is smaller.
In contrast, for $U=0$ and intermediate 
concentrations,  the dephasing rate displays a dip at the 
$M$ point and is maximum close to the $X$ point (see e.g. Fig.\ref{Fig3}). 
The  double--peak momentum dependence of  $\Gamma_{{\bf Q}}$ 
for $U$=0
can be recovered for large $U$ only  by introducing 
a sufficiently large itinerant spin 
damping rate $\Gamma$ (see  
Fig.\ref{Fig6}(c)).
The experiment of Ref.[\onlinecite{ye}] 
observed an 
increase in the spin--wave damping along 
$\Gamma \to M$ that exceeds the corresponding 
increase along  $\Gamma \to X$, 
similar to our results for finite $U$ and  $\Gamma$=0
(Fig.\ref{Fig6}(b)).  
Our calculations 
show that such behavior of the spin damping 
can be attributed to 
the  Coulomb repulsion $U$ in the intrinsic system 
($\Gamma,\gamma \rightarrow$0) described by the Hamiltonian 
$H$.

\begin{figure}[t]
\vspace{0.38 in}
\centerline{
\hbox{\psfig{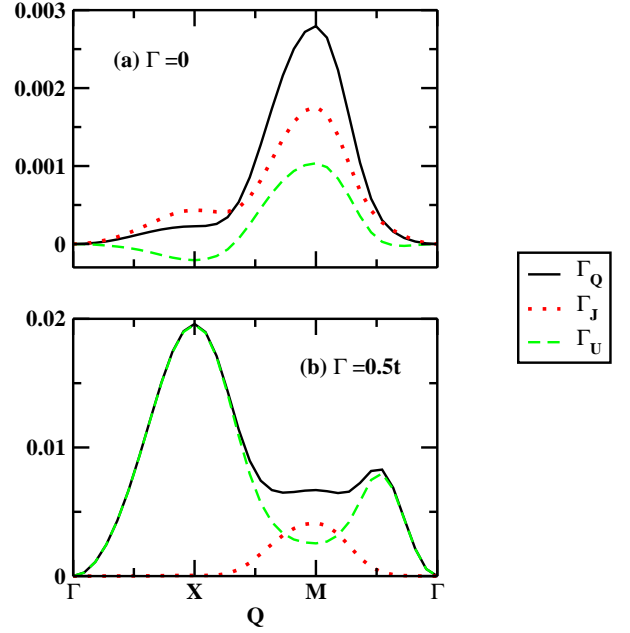}}}
\caption{(Color online) 
Contributions $\Gamma_{{\bf Q}}^J$ and $\Gamma_{{\bf Q}}^U$
 to the inelastic 
spin--wave dephasing rate $\Gamma_{{\bf Q}}$. 
(a) $\Gamma=0$, (b)  $\Gamma$=0.5t. $n$=0.6, $J$=2$t$, $U$=10$t$, 
$\gamma$=0.2$t$, $J_{AF}=0$. }\label{Fig7}
\end{figure}

Fig.\ref{Fig6} also demonstrates a 
qualitative difference in the dependence of 
$\Gamma_{{\bf Q}}$ on $U$ between the cases of  small 
and large itinerant spin damping. 
For $\Gamma$=0, 
the spin--wave  dephasing rate  {\em increases}
with $U$ (Fig.\ref{Fig6}(b)), while 
for $\Gamma$=0.5$t$ it {\em decreases} 
with $U$ close to the $X$--point  
(Fig.\ref{Fig6}(c)).
This result indicates that the  magnetization relaxation 
may depend  on the interplay between  Coulomb repulsion and 
the dephasing of the  itinerant spin 
via the coupling to an external bath.

To interpret the above behaviors, we plot in Fig.\ref{Fig7} the two
contributions to $\Gamma_{{\bf Q}}$ obtained from the 
self energies  Eqs.(\ref{self-J}) and 
(\ref{self-U}) for zero and finite $\Gamma$. 
 $\Gamma^J$ is determined by $G$, 
while $\Gamma^U$ is determined by 
$\Phi$. 
For $\Gamma$=0, Fig.\ref{Fig7}(a) shows that 
$\Gamma^J$ and $\Gamma^U$  are comparable in magnitude, 
since in this case they both arise from $\rm{Im}$G. 
On the other hand,  as 
$\Gamma$ increases, 
the relative magnitude of
$\Gamma^J$ and $\Gamma^U$ changes and the latter  dominates
(see Fig.\ref{Fig7}(b)).
This enhancement of 
$\Gamma^U$ arises from the
additional contribution to 
$\rm{Im}\Phi$, 
Eq.(\ref{Phi-F}), 
obtained by adding 
the relaxation rate $\Gamma$
to 
Eq.(\ref{Phi-F}).
Even though 
$\Gamma^J$ continues to have the same momentum dependence as 
for $\Gamma$=0, 
 $\Gamma^U$
does  not 
(compare  Figs.\ref{Fig7}(a) and \ref{Fig7}(b)).

\begin{figure}[t]
\vspace{0.38 in}
\centerline{
\hbox{\psfig{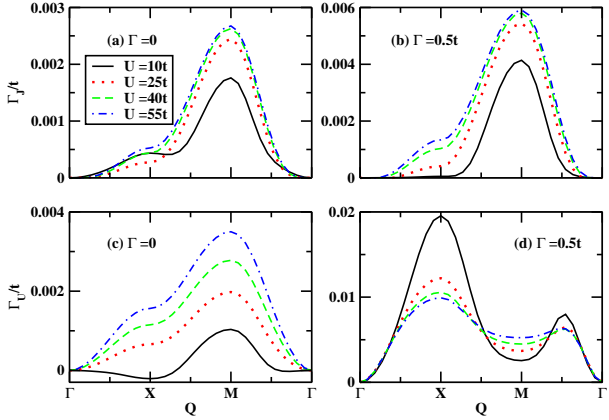}}}
\caption{(Color online) Dependence of 
$\Gamma_{{\bf Q}}^J$ and $\Gamma_{{\bf Q}}^U$
on  Coulomb repulsion for large $U$. 
$n$=0.6, $J$=2$t$, 
$\gamma$=0.2$t$, $J_{AF}=0$.}\label{Fig8}
\end{figure}

Fig.\ref{Fig8} shows 
the dependence of the two contributions 
to the inelastic 
dephasing rate on the Coulomb repulsion 
for large $U$ and different $\Gamma$.   
 $\Gamma^J$ increases and then saturates with increasing $U$.
For 
$\Gamma$=0 (intrinsic system),   
 $\Gamma^U$ increases with $U$ and eventually exceeds 
$\Gamma^J$. 
Unlike for $\Gamma^J$,
the dependence 
of $\Gamma^U$ on the momentum and  on $U$ 
is qualitatively
different for large and small 
$\Gamma$. 
For example, 
 $\Gamma^U$ 
decreases with $U$ at the $X$--point 
for  large $\Gamma$ but increases 
for $\Gamma$=0. 
The above behavior of $\Gamma^U$ 
dominates the total dephasing rate
for large $\Gamma$.

\begin{figure}[t]
\vspace{0.38 in}
\centerline{
\hbox{\psfig{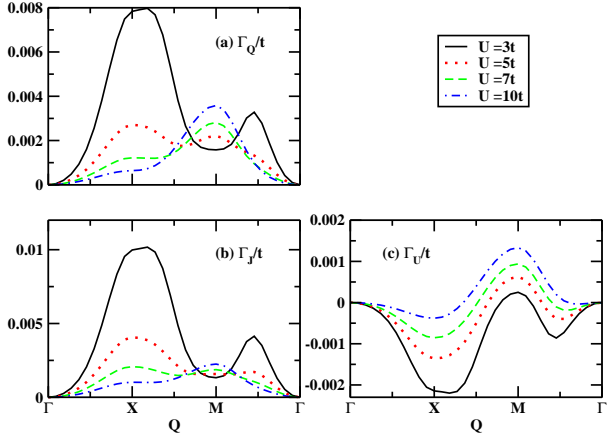}}}
\caption{(Color online) Dependence of 
(a) $\Gamma_{{\bf Q}}$, (b) $\Gamma_{{\bf Q}}^J$, 
(c) $\Gamma_{{\bf Q}}^U$ 
on the Coulomb repulsion for small $U$. 
$n$=0.6, $J$=4$t$, 
$\gamma$=0.2$t$, $J_{AF}=\Gamma=0$.}\label{Fig9}
\end{figure}

Fig.\ref{Fig9}(a) shows the transition in the momentum dependence of
$\Gamma_{{\bf Q}}$ 
as  $U$ increases for $\Gamma$=0.
This transition 
occurs around $U \sim$6$t$, where the double peak momentum 
dependence for $U$=0, 
with a dip at the $M$--point,  
changes into a  peak at the $M$--point. 
As can be seen in Fig.\ref{Fig9}(b), the above transition 
arises from the changes in the behavior of 
$\Gamma^J$, determined by the Green's function $G$ (Eq.(\ref{G-F})), 
that are
introduced by the Coulomb repulsion.
Fig.\ref{Fig9}(c), on the other
hand, shows that the overall momentum dependence of 
$\Gamma^U$ remains approximately the same for all $U$.

\begin{figure}[t]
\vspace{0.38 in}
\centerline{
\hbox{\psfig{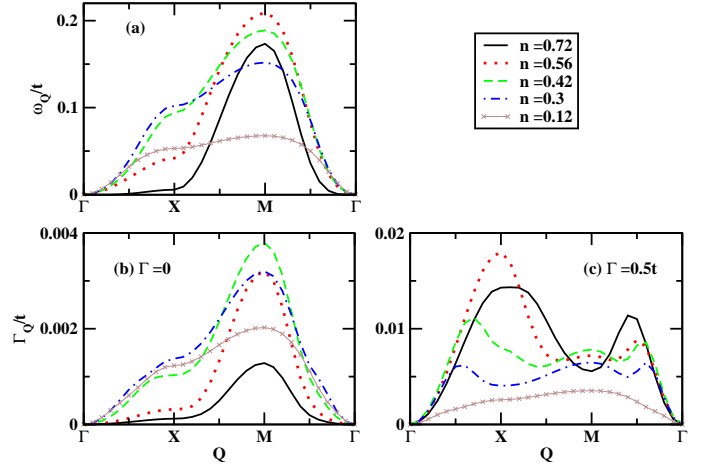}}}
\caption{(Color online) Dependence of spin--wave dispersion 
 and 
dephasing rate on the carrier concentration $n$ for 
$U$=10$t$. 
(a) dispersion, 
(b) dephasing  rate for $\Gamma$=0, 
(c) dephasing  rate for $\Gamma$=0.5$t$. 
$J =2t, J_{\rm{AF}}=0, \gamma =0.2t$.}\label{Fig10}
\end{figure}

Finally, we turn to the possibility 
of controlling the magnetization relaxation
by tuning the carrier concentration $n$ 
and study how the  Hubbard repulsion $U$ changes the picture 
as compared to the prediction of 
Fig.\ref{Fig4}.
Fig.\ref{Fig10} shows the dependence of the  dispersion 
and spin damping rate 
on $n$ 
 within a wide  range  of concentrations, $n=0.7-0.1$. 
As can be seen in Fig.\ref{Fig10}(a), 
the pronounced softening along the $\Gamma$--$X$ 
direction
disappears rapidly with decreasing $n$ (or increasing hole doping 
$x$=1-$n$). For small values of $n$, the overall energies 
decrease  and the overall shape of the dispersion changes.

The above $n$--dependence of the spin--wave dispersion 
correlates with corresponding changes in the  dephasing rate. 
Fig.\ref{Fig10}(b), obtained for $\Gamma$=0, 
shows that the  Coulomb repulsion $U$ changes drastically  the 
dependence of $\Gamma_{{\bf Q}}$ on $n$ 
as compared to the predictions of the minimal double exchange model, 
Fig.\ref{Fig4}(a). 
In this case, 
the spin dephasing rate {\em increases}
as $n$ decreases to intermediate values, in a way correlated with the 
disappearance of the spin--wave softening and non--Heisenberg 
behavior. For small $n$, where the softening has disappeared and 
the overall energies start to decrease, 
the spin damping rate also decreases.  
Furthermore, unlike for 
$U$=0 or for large $\Gamma$, 
$\Gamma_{{\bf Q}}$ 
displays a strong {\em maximum} at point $M$
{\em for all concentrations} 
and is always weaker along the $\Gamma$--$X$ direction.
As can be seen by comparing 
Figs.\ref{Fig10}(a) and (b), a sufficiently short itinerant spin lifetime 
(large $\Gamma$) 
changes drastically the momentum 
dependence of $\Gamma_{{\bf Q}}$ 
and its dependence on $n$.

We conclude based on 
Fig.\ref{Fig10} 
that the magnitude and momentum dependence of 
$\Gamma_{{\bf Q}}$, as well as the 
spin--wave softening,  can be controlled  by tuning the 
carrier  concentration $n$, via hole doping or by external means 
such as  photoexcitation. 
\cite{bigot-05,kampen,kimel-anis,tolk,lupke,kimel-rev,wang-rev,shah,chovan}
It is  clear that 
the on--site Coulomb (Hubbard) repulsion
plays a dominant role,
by inducing new  correlations and dynamics 
absent in the simple double--exchange  model. Such correlations  change 
drastically the momentum dependence 
and magnitude of $\Gamma_{{\bf Q}}$ with varying $n$
and  must be treated in a consistent way in order to
arrive at trustworthy conclusions and comparisons 
to experiment.  
Our results suggest that, as a first step,  a systematic
experimental study of the magnetization dynamics 
as function of doping $x$ and a comparison to 
the theory
is necessary in order to  decide which many--body mechanisms 
dominate the collective magnetization dynamics and relaxation 
and learn how to control this dynamics for potential 
magnetic device and spintronics 
applications. 
     
\section{Conclusions}
\label{concl}

In this paper we presented a general method 
for describing the spin--wave dynamics and relaxation 
in itinerant ferromagnets. 
This method is based on a correlation expansion 
of the Green's function equations of motion 
that systematically treats all correlations between any given number of
elementary excitations. 
Using this method,
we derived a closed system of equations
that treat the magnetic exchange and Coulomb interactions 
non--perturbatively and 
solved it to obtain the 
Green's function that determines the transverse spin susceptibility. 
Our results for the 
spin--wave dispersion reproduce 
previous variational \cite{kapet1,kapet2} and 
exact diagonalization \cite{exact,igar}  
results 
(in the limit $\Gamma$,$\gamma$ $\rightarrow$0) 
and therefore allow us to draw definite conclusions regarding 
the magnitude of the spin--wave softening.
Using the properties of the 
fully polarized Hartree--Fock ground state 
with maximum spin, 
we showed that our method gives 
the  exact  spin Green's function within a subspace of 
states that include up to one Fermi sea pair excitation.
Our factorization scheme of the higher Green's functions 
also applies to other ground states. 
Our results recover the 1/S expansion 
results as special case.  
We showed that, in the parameter regime of 
interest in the manganites, 
 the latter  approximation overestimates the 
spin--wave softening, while at the same time 
it grossly underestimates the spin--wave damping rate. 
Furthermore, by comparing with the ladder 
approximation treatment of the vertex corrections
to the magnon--carrier scattering, 
which treats the multiple magnon--electron 
scatterings while neglecting the interactions with a Fermi sea hole,  
we showed that three--body correlations 
between the magnon and an electron--hole Fermi sea pair excitation have 
 an important effect on the spin relaxation.

Using the above many--body theory, we 
calculated the inelastic spin--wave dephasing rate non--perturbatively 
in the interactions and 1/S (i.e. beyond the standard Fermi's Golden Rule). 
We showed that  correlations between a carrier 
spin--flip excitation and a Fermi sea pair induced by 
the Coulomb repulsion $U$ play a very important role 
in the parameter regime relevant to the manganites. 
We also showed that both the magnitude and  momentum dependence of the 
spin--wave dephasing rate depend sensitively on the itinerant 
carrier concentration. This result implies the possibility of 
controlling the magnetization relaxation in itinerant ferromagnets by 
tuning the carrier concentration, either via doping or 
by external means (e.g. photoexcitation or by using electric fields 
and currents or gates). We also argued that 
the interplay between  on--site 
Coulomb (Hubbard) interaction  
and a finite 
itinerant carrier spin lifetime, due to the coupling of 
the different spin states 
induced by spin--orbit or other interactions not included in our 
Hamiltonian, can 
affect our results in the realistic system. 
The momentum dependence of the spin--wave dephasing rate 
observed in recent experiments \cite{ye} is consistent 
with the results of our calculation only for sufficiently large 
$U$ and  $\Gamma \rightarrow$0. 
In all other cases, we obtain a distinctly different 
double peak momentum dependence.
Although the  bandstructure of the relevant materials 
must be included in order to arrive at quantitative comparisons 
with the experiment, our calculation already demonstrates the crucial role 
of 
correlations.
The agreement of the main trends 
in  spin--wave softening 
 and  damping rate changes 
as function of $n$ between our theory and the experiment 
suggests that 
the simple one--band model 
already contains the main inelastic scattering processes 
and correlations.
Our results suggest that
ultrafast magneto--optical pump--probe spectroscopy experiments, 
which directly probe the changes in spin relaxation  
and dynamics induced by  photoexcited carriers,\cite{chovan}  
may provide new insight into the physics of the manganites and 
other itinerant ferromagnetic systems.

This work was supported by the EU STREP program HYSWITCH.

\end{document}